\documentclass[prl,twocolumn,superscriptaddress,showpacs]{revtex4}

\usepackage{graphics}           
\usepackage{bm}                 
\usepackage{amsmath}            
\usepackage{amssymb}
\usepackage{verbatim}           
\usepackage{amsthm}             
\theoremstyle{plain}            

\def\bra#1{{\langle#1|}}
\def\ket#1{{|#1\rangle}}
\def\bracket#1#2{{\langle#1|#2\rangle}}

\def\expect#1{{\langle#1\rangle}}
\def\e{{\rm e}}
\def\proj{{\hat{\cal P}}}
\def\tr{{\rm Tr}}
\def\H{{\hat H}}

\def\Lop{{\cal L}}
\def\Ahat{{\hat A}}
\def\Adag{{\hat A}^\dagger}
\def\Ehat{{\hat E}}

\def\Shat{{\hat S}}
\def\Sdag{{\hat S}^\dagger}
\def\U{{\hat U}}
\def\Udag{{\hat U}^\dagger}
\def\Zhat{{\hat Z}}

\def\Op{{\hat O}}
\def\id{{\hat I}}
\def\Pr{{\proj_{\rm R}}}
\def\Pl{{\proj_{\rm L}}}
\def\x{{\hat x}}

\begin{document}

\title{The quantum to classical transition for random walks}

\author{Todd A. Brun}
\email{tbrun@ias.edu}
\affiliation{Institute for Advanced Study, Einstein Drive, Princeton, 
              NJ 08540}

\author{Hilary A. Carteret}
\email{hcartere@cacr.math.uwaterloo.ca}
\affiliation{Department of Combinatorics and Optimization, 
             University of Waterloo, Waterloo, Ontario, N2L 3G1, Canada}

\author{Andris Ambainis}
\email{ambainis@ias.edu}
\affiliation{Institute for Advanced Study, Einstein Drive, Princeton, 
              NJ 08540}

\date{August 2002}

\begin{abstract}
We look at two possible routes to classical behavior for the discrete
quantum random walk on the line:  decoherence in the quantum
``coin'' which drives the walk, or the use of higher-dimensional
coins to dilute the effects of interference.  We use the
position variance as an indicator of classical behavior,
and find analytical expressions for this in the long-time
limit; we see that the multicoin walk retains the ``quantum''
quadratic growth of the variance except in the limit of
a new coin for every step, while the walk with decoherence exhibits
``classical'' linear growth of the variance even for weak decoherence.
\end{abstract}

\pacs{05.40.Fb 03.65.Ta 03.67.Lx}

\maketitle

\section{Introduction}

Considerable work has been done recently on quantum random walks,
which are unitary (and hence reversible) systems
designed as analogues to the usual classical case.  Two approaches
have been taken to the problem:  {\it continuous} \cite{Continuous}
and {\it discrete} \cite{Discrete,Meyer96,NayaknV,Decoherence}
walks.  This paper is exclusively concerned with the discrete walk
on the infinite line.
In this case, we introduce an extra ``coin'' degree of freedom into
the system.  As in the classical random walk, the outcome of a
``coin flip'' determines the direction that the particle moves.
In the quantum case, however,
both the flip of the coin and the conditional motion of the particle
are unitary transformations.  Different possible classical paths
can therefore interfere.

For the classical walk, $p(x,t)$ has the form of a
binomial distribution, with a variance ${\bar{x^2}}-{\bar x}^2$ which
grows linearly with time.  The variance in the quantum walk, by contrast,
grows {\it quadratically} with time \cite{Meyer96}; and the
distribution $p(x,t)$ has a complicated, oscillatory form
\cite{NayaknV}.  Both of these are effects of interference between the
possible paths of the particle.  There are two obvious ways to regain
the classical behavior.  If the quantum coin is measured at every step,
then the record of the measurement outcomes singles out a particular
classical path.  By averaging over all possible measurement records,
one recovers the usual classical behavior.

Alternatively, rather than measuring the coin every time, one could
replace it with a {\it new} quantum coin for each flip.  After a time
$t$ one would have accumulated $t$ coins, all of them entangled with
the position of the particle.  By measuring them, one could reconstruct
an unique classical path; averaging over the outcomes would once again
produce the classical result.

These two approaches, which are equivalent in the classical limit,
give two different routes from quantum to classical.
We might increase the number of coins used to generate the walk, cycling
among $M$ different coins, in the limit using a new coin at each step.
Or we might {\it weakly} measure the coin after each step, reaching the
classical limit with strong, projective measurements.  This is equivalent
to having a coin which is subject to {\it decoherence}.

In this paper we contrast these two approaches, using the functional
dependence of the variance on time as an indicator of ``classical'' vs.
``quantum'' behavior.  In the presence even of very weak decoherence,
the variance of the quantum walk grows linearly with $t$ at long times;
while even using a large number of coins, the the variance of the
unitary walk grows quadratically.

Let us consider a fairly general quantum walk on the line.
The particle degree of freedom has a basis of position eigenstates
$\{\ket{x}\}$, $\x\ket{x} = x\ket{x}$, where $x$ can be any integer.
We will assume that the
particle begins the walk at the origin, in state $\ket0$.  The walk
is driven by a separate
``coin'' degree of freedom:  a $D$-dimensional
system with an initial state $\ket{\Phi_0}$.  Let $\Pr,\Pl$ be two
orthogonal projectors on the Hilbert space of the ``coin,'' such that
$\Pr + \Pl = \id$.  These represent the two possible outcomes of
the coin flip,  Right or Left.  We assume that the coin is unbiased,
meaning $\tr\Pr = \tr\Pl = D/2$.
We also define a unitary transformation $\U$
which ``flips'' the coin.  One step of the quantum walk is
given by the unitary operator
\begin{equation}
\Ehat \equiv \left(\Shat \otimes \Pr + \Sdag \otimes \Pl\right)
  \left( \id \otimes \U \right) \;,
\end{equation}
where $\Shat,\Sdag$ are shift operators on the particle position,
$\Shat\ket{x} = \ket{x+1}$, $\Sdag\ket{x} = \ket{x-1}$.
The full initial state of the system (particle and coin) is
$\ket{\Psi_0} = \ket0 \otimes \ket{\Phi_0}$.

We can identify the eigenvectors $\ket{k}$ of $\Shat,\Sdag$,
\begin{equation}
\ket{x} = \int_{-\pi}^{\pi} \frac{dk}{2\pi} \e^{-ikx}\ket{k} \;,
\end{equation}
with eigenvalues
\begin{equation}
\Shat\ket{k} = \e^{-ik}\ket{k} \;, \ \ 
\Sdag\ket{k} = \e^{+ik}\ket{k} \;.
\end{equation}
In the $k$ basis, the evolution operator becomes
\begin{eqnarray}
\Ehat\left( \ket{k} \otimes \ket\Phi \right)
  &=& \ket{k} \otimes \left( \e^{-ik} \Pr + \e^{ik} \Pl\right) \U
  \ket\Phi \;, \nonumber\\
  &\equiv& \ket{k} \otimes \U_k \ket\Phi \;,
\end{eqnarray}
where $\U_k$ is also a unitary operator.

We now generalize to allow for decoherence.  Suppose
that before each unitary ``flip'' of the coin, a {\it completely positive}
and {\it unital} map is performed on the coin.
This map is given by a set of operators $\{\Ahat_n\}$
on the coin degree of freedom which satisfy
\begin{equation}
\sum_n \Adag_n\Ahat_n =
\sum_n \Ahat_n\Adag_n = \id \;.
\end{equation}
A density operator $\chi$ for the coin degree of freedom is transformed
\begin{equation}
\chi \rightarrow \chi' = \sum_n \Ahat_n \chi \Adag_n \;.
\end{equation}
We combine this with the unitary evolution to define an
evolution {\it superoperator}
\begin{equation}
\Lop_{kk'} \chi \equiv \sum_n
  \U_k \Ahat_n \chi \Adag_n \Udag_{k'} \;.
\end{equation}
Note that for the diagonal case $k=k'$, this superoperator
is also unital, and hence preserves the identity.

The initial state is
\begin{equation}
\rho_0 = \ket{\Psi_0}\bra{\Psi_0} =
  \int \frac{dk}{2\pi}\int \frac{dk'}{2\pi} \ket{k}\bra{k'} \otimes
  \ket{\Phi_0}\bra{\Phi_0} \;.
\end{equation}
Let the quantum random walk proceed for $t$ steps.  Then the state
evolves to
\begin{equation}
\rho_t =
  \int \frac{dk}{2\pi}\int \frac{dk'}{2\pi} \ket{k}\bra{k'} \otimes
  \Lop_{kk'}^t  \ket{\Phi_0}\bra{\Phi_0} \;.
\end{equation}
From this, the probability to reach a point $x$ at time $t$ is
\begin{eqnarray}
p(x,t) &=&
  \int \frac{dk}{2\pi} \int \frac{dk'}{2\pi} \bracket{x}{k}\bracket{k'}{x}
  \tr\left\{ \Lop_{kk'}^t \ket{\Phi_0}\bra{\Phi_0} \right\}
  \nonumber\\
&=& \int \frac{dk}{2\pi} \int \frac{dk'}{2\pi} \e^{-ix(k-k')}
  \tr\left\{ \Lop_{kk'}^t \ket{\Phi_0}\bra{\Phi_0} \right\} \;.
\label{probdistribution}
\end{eqnarray}

Eq.~(\ref{probdistribution}) for $p(x,t)$ will be difficult to evaluate
analytically; hence, we restrict
our interest to the {\it moments} of this distribution.
\begin{eqnarray}
\expect{\x^m}_t &=& \sum_x x^m p(x,t) \nonumber\\ 
&=& \sum_x x^m \int \frac{dk}{2\pi}
  \int \frac{dk'}{2\pi} \e^{-ix(k-k')} \nonumber\\
&& \times \tr\left\{ \Lop_{kk'}^t \ket{\Phi_0}\bra{\Phi_0} \right\} \;.
\end{eqnarray}
We invert the order of operations and do the $x$ sum first, yielding
\begin{eqnarray}
\expect{\x^m}_t &=&
  \frac{(-i)^m}{2\pi} \int dk \int dk' \delta^{(m)}(k-k') \nonumber\\
&& \times \tr\left\{ \Lop_{kk'}^t \ket{\Phi_0}\bra{\Phi_0} \right\} \;,
\label{momentform}
\end{eqnarray}
where $\delta^{(m)}(k-k')$ is the $m$th derivative of the delta
function.  We can then integrate this by parts.

In integrating (\ref{momentform}) we will need
\begin{eqnarray}
\frac{d}{dk} \tr\left\{ \Lop_{kk'} \Op \right\}
  &=& - i \tr\left\{ \Zhat \Lop_{kk'} \Op \right\} \;, \nonumber\\
&=& - i \tr\left\{ (\Lop_{kk'} \Op) \Zhat \right\} \;, \nonumber\\
&=& - \frac{d}{dk'} \tr\left\{ \Lop_{kk'} \Op \right\} \;, \nonumber\\
\Zhat &\equiv& \Pr - \Pl \;.
\label{lopderiv}
\end{eqnarray}
Making use of (\ref{lopderiv}), when we
carry out the integration by parts for the first moment we get
\begin{equation}
\expect{\x}_t = - \int \frac{dk}{2\pi} \sum_{j=1}^t
  \tr\left\{ \Zhat \Lop_{k}^j \ket{\Phi_0}\bra{\Phi_0} \right\} \;,
\label{firstmoment1}
\end{equation}
where we've introduced the simplified notation
$\Lop_{k} \equiv \Lop_{kk}$.

We can carry out a similar integration by parts to get the second moment:
\begin{eqnarray}
\expect{\x^2}_t &=& \int \frac{dk}{2\pi} \Biggl[
 \sum_{j=1}^t \sum_{j'=1}^j 
  \tr\left\{ \Zhat \Lop_k^{j-j'} ( \Zhat \Lop_k^{j'}
  \ket{\Phi_0}\bra{\Phi_0} ) \right\} \nonumber\\
&& + \sum_{j=1}^t \sum_{j'=1}^{j-1} 
  \tr\left\{ \Zhat \Lop_k^{j-j'} ( ( \Lop_k^{j'}
  \ket{\Phi_0}\bra{\Phi_0} ) \Zhat ) \right\} \Biggr] \;.
\label{secondmoment1}
\end{eqnarray}

Let us for the moment specialize to the unitary case, so
$\Lop_k\rho = \U_k \rho \Udag_k$.  In this case, we can expand
$\ket{\Phi_0}$ in terms of the eigenvectors of $\U_k$:
\begin{equation}
\ket{\Phi_0} = \sum_l c_{kl} \ket{\phi_{kl}} \;,\ \ 
  \U_k \ket{\phi_{kl}} = \e^{i\theta_{kl}} \ket{\phi_{kl}} \;.
\end{equation}
Assume, for the moment, that $\U_k$ is nondegenerate, so the
$\theta_{kl}$ are all distinct.
If we plug these expressions into (\ref{firstmoment1}) and
(\ref{secondmoment1}), we notice that most of the terms will be
{\it oscillatory}; at long times $t$, they will average to zero.
The only terms that survive will be diagonal in $l$:
\begin{eqnarray}
\expect{\x}_t &=& - \sum_{j=1}^t \int \frac{dk}{2\pi}
   \bra{\Phi_0} (\U_k)^j \Zhat (\Udag_k)^j \ket{\Phi_0} \nonumber\\
&=& - t \int \frac{dk}{2\pi} \sum_{l=1}^D |c_{kl}|^2
   \bra{\phi_{kl}} \Zhat \ket{\phi_{kl}} \nonumber\\
&& + {\rm oscillatory\ terms} \;.
\label{firstmoment_unitary}
\end{eqnarray}

Similarly, for the second moment
\begin{eqnarray}
\expect{\x^2}_t
&=&  t^2 \int \frac{dk}{2\pi} \sum_{l=1}^D |c_{kl}|^2
   \bra{\phi_{kl}} \Zhat \ket{\phi_{kl}}^2 \nonumber\\
&&  + O(t) + {\rm oscillatory\ terms} \;.
\label{secondmoment_unitary}
\end{eqnarray}
So in the long-time limit, the variance will always grow {\it quadratically}
in time for a unitary coin of finite dimension.
If $\U_k$ is degenerate, the expressions (\ref{firstmoment_unitary})
and (\ref{secondmoment_unitary}) will have to be modified to include
appropriate cross-terms; but this will not change the functional dependence
on $t$.

The usual case considered in the literature has taken the coin
to be a simple two-level system, and the ``flip'' operator $\U$
to be the usual Hadamard transformation $\H$:
\begin{eqnarray}
\H \ket{R} &=& \frac{1}{\sqrt{2}} \left( \ket{R} + \ket{L} \right) \;, 
  \nonumber\\
\H \ket{L} &=& \frac{1}{\sqrt{2}} \left( \ket{R} - \ket{L} \right) \;.
\end{eqnarray}
The projectors are $\Pr = \ket{R}\bra{R}$, $\Pl = \ket{L}\bra{L}$.
The walk on the line in this case has been exactly solved by
Nayak and Vishwanath \cite{NayaknV}, and agrees with the expression
(\ref{secondmoment_unitary}) given above.  We have also considered
the case of a walk driven by $M$ coins, flipped cyclically \cite{Multicoin}.
In this case, (\ref{firstmoment_unitary}) and
(\ref{secondmoment_unitary}) can be solved analytically; at long times,
the variance goes like
\begin{equation}
\expect{\x^2}-\expect{\x}^2 = t^2 \left( \frac{3-\sqrt{8}+1/M}{\sqrt{32}}
  \right) + O(t) + {\rm osc.\ terms} \;.
\label{multicoin_variance}
\end{equation}
We compare this to the results of numerical simulations in figure 1,
finding the agreement to be excellent.  Note that the quadratic
growth of the variance doesn't vanish even in the limit of large
$M$.  Only with a new coin for every step ($M=t$) do we recover the
classical behavior.  (Note that this doesn't contradict the result
(\ref{multicoin_variance}), since that is only strictly valid for $t \gg M$.)

\begin{figure}[t]
\includegraphics{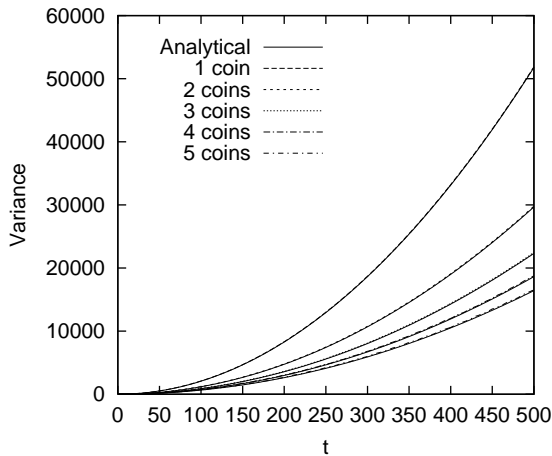}
\caption{\label{fig1}  The variance $\expect{\x^2}-\expect{\x}^2$
for the unitary walk with multiple coins, $M$=1--5.
The coins all begin in state $\ket{R}$.
The variance grows quadratically in time, in contrast to the linear
growth in the classical random walk.}
\end{figure}

Let's now allow for decoherence.  Because the superoperator
$\Lop_k$ is unital it preserves the identity $\Lop_k\id = \id$;
its largest eigenvalue is 1.  We explicitly assume that all
the other eigenvalues of $\Lop_k$ obey $|\lambda| < 1$.  It is handy
then to separate out the traceless part of the initial state
\begin{equation}
\rho_0
  = \id/D + (\ket{\Phi_0}\bra{\Phi_0} - \id/D)
  \equiv \id/D + \chi_0 \;.
\end{equation}

Plugging this into the equation (\ref{firstmoment1}) we get
\begin{eqnarray}
\expect{\x}_t &=& - \sum_{j=1}^t \int \frac{dk}{2\pi}
  \tr\left\{ \Zhat/D \right\} +
  \tr\left\{ \Zhat \Lop_{k}^j \chi_0 \right\} \nonumber\\
&=& - t \tr\left\{ \Zhat/D \right\} \nonumber\\
&&  - \int \frac{dk}{2\pi} \tr\left\{ 
  \Zhat (1-\Lop_k)^{-1} (\Lop_k - \Lop_k^{t+1}) \chi_0 \right\} \;.
\end{eqnarray}
Note that $(1-\Lop_k)^{-1}$ acts only on the traceless
operator $\chi_0$, so it is well-defined.
If the coin is unbiased (as we assumed), then $\tr\Zhat=0$ and
the first term vanishes.  At long times the
$\Lop_k^{t+1}$ term will decay away.  So in the long time limit,
the first moment of the walk with a decoherent coin will tend
to a constant.

The second moment is more complicated in detail, but similar in spirit.
Again separating out the traceless part $\chi_0$, we get
\begin{eqnarray}
\expect{\x^2}_t &=& \int \frac{dk}{2\pi} \sum_{j=1}^t \Biggl[
 \tr\left\{ \Zhat^2 \Lop_k^j \rho_0 \right\} \nonumber\\
&+& \sum_{j'=j+1}^t \Biggl(
  \tr\left\{ \Zhat \Lop_k^{j'-j} ( 2 \Zhat/D ) \right\} \nonumber\\
&& + \tr\left\{ \Zhat \Lop_k^{j'-j} ( \Zhat (\Lop_k^{j} \chi_0) +
  ( \Lop_k^{j} \chi_0 ) \Zhat ) \right\} \Biggr) \Biggr] \nonumber\\
&=& \int \frac{dk}{2\pi} \sum_{j=1}^t \Biggl[ 1 +
  (2/D) \tr\biggl\{ \Zhat (1-\Lop_k)^{-1} \nonumber\\
&& \times (\Lop_k - \Lop_k^{t-j+1}) \Zhat \biggr\} \nonumber\\
&& + \tr\biggl\{ \Zhat (1-\Lop_k)^{-1} (\Lop_k - \Lop_k^{t-j+1}) \nonumber\\
&& \times  ( \Zhat (\Lop_k^{j} \chi_0) +
    ( \Lop_k^{j} \chi_0 ) \Zhat ) \biggr\} \Biggr] \;.
\end{eqnarray}
It is not difficult to see that the last term will tend towards a
constant for large $t$, while the first two will grow linearly.
So we get the approximate expression at long times
\begin{eqnarray}
\expect{\x^2}_t &\approx& t \left( 1 + \frac{1}{\pi D} \int dk
  \tr\left\{ \Zhat (1-\Lop_k)^{-1}
    \Lop_k \Zhat \right\} \right) \nonumber\\
&& + {\rm const.} \;,
\label{secondmoment_decoherent}
\end{eqnarray}
which is linear in $t$.

If we specialize to the case of a single two-level coin undergoing
the Hadamard evolution, we can find exact solutions to
(\ref{secondmoment_decoherent}).  
We first pick a particular form for the decoherence.  The most
convenient numerically is {\it pure dephasing}
\begin{equation}
\Ahat_{0,1} = \frac{1}{\sqrt2} \left( \e^{\pm i\theta} \ket{R}\bra{R} +
  \e^{\mp i\theta} \ket{L}\bra{L} \right) \;.
\label{dephasing}
\end{equation}
For $\theta=0$ this reverts to the unitary case, while for
$\theta=\pi/4$ there is complete decoherence at each step.  At
long times we have \cite{DecoherentCoin}
\begin{equation}
\expect{\x^2}_t - \expect{\x}^2_t \approx
  t \left( \cot^2 2\theta + \csc^2 2\theta \right)
  + {\rm const.} \;,
\end{equation}
which goes to 1 as $\theta \rightarrow \pi/4$, and diverges as
$\theta \rightarrow 0$, when the long-time approximation breaks down.
In figure 2, we compare this result to the output of numerical simulations,
once more finding excellent agreement.  Note that while the variance
grows linearly, as in the classical case, it grows {\it faster} than
the classical case.  The reflects the persistent effect of interference,
which causes the particle to continue to drift in a particular direction.

\begin{figure}[t]
\includegraphics{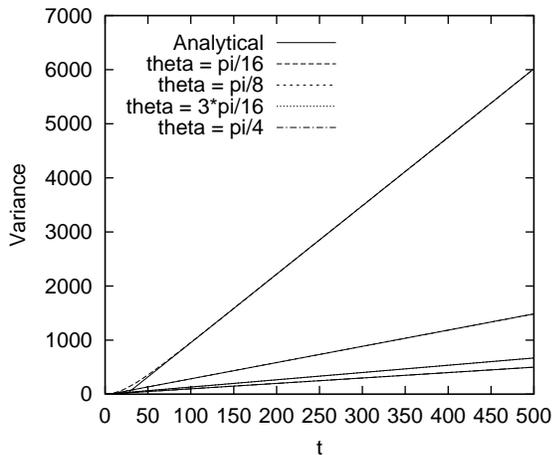}
\caption{\label{fig2}  $\expect{\x^2}-\expect{\x}^2$ vs. $t$ for the
quantum random walk with decoherence, for $\theta=\pi/16,\pi/8,3\pi/16,\pi/4$.
For all cases the coin began in the initial
state $\ket{R}$.  Note that the variance goes asymptotically to a
linear growth at long times which matches our analytical estimate;
the rate of growth goes to one with increasing decoherence, matching
the classical case at $\theta=\pi/4$.}
\end{figure}

From the time-dependence of the moments, one
might reasonably claim that the multicoin system remains ``quantum''
even in the limit of very large numbers of coins, while the
decoherent system remains ``classical'' even in the limit of very
weak noise.  Moreover, from equations (\ref{secondmoment_unitary})
and (\ref{secondmoment_decoherent}) this behavior seems to be generic,
independent of the particular model chosen.

We should emphasize that in this paper we have only altered the
coin degree of freedom.  One might naturally consider modifications
of the evolution of the particle as well, such as allowing decoherence
of the position as well as the coin.  There have been numerical
studies of this, and of the transition to classical behavior that
results \cite{Decoherence}.

\begin{acknowledgments}

We would like to thank Bob Griffiths, Lane Hughston, Viv Kendon,
and Michele Mosca for useful conversations.
TAB acknowledges financial support from
the Martin A.~and Helen Chooljian Membership in Natural Sciences,
and DOE Grant No.~DE-FG02-90ER40542.  AA was supported by NSF grant
CCR-9987845, and by the State of New Jersey.  HAC was supported by
MITACS, The Fields Institute, and NSERC CRO project ``Quantum
Information and Algorithms.''

\end{acknowledgments}


\end{document}